\documentclass[11pt,dvips]{article}
\usepackage{graphicx}
\usepackage{amsmath}
\usepackage{amssymb}
\usepackage{latexsym}
\usepackage{color}
\begin{document}
\thispagestyle{empty}
\begin{center}

\vspace{1.8cm}

 {\large {\bf Hilbert-Schmidt measure of pairwise quantum  discord for three-qubit $X$ states }}\\

\vspace{1.5cm} {\bf M. Daoud}$^{a,b,c}${\footnote { email: {\sf
m$_{-}$daoud@hotmail.com}}}, {\bf R. Ahl Laamara}$^{d,e}$ {\footnote
{ email: {\sf ahllaamara@gmail.com}}}  and {\bf S. Seddik}$^{d}$
{\footnote { email: {\sf sanaa1.sanaa@hotmail.fr}}} \\
\vspace{0.5cm}
$^{a}${\it Max Planck Institute for the Physics of Complex Systems, Dresden, Germany}\\[0.5em]
$^{b}${\it Abdus Salam  International Centre for Theoretical Physics, Trieste, Italy}\\[0.5em]
$^{c}${\it Department of Physics, Faculty of Sciences,  University Ibnou Zohr, Agadir , Morocco}\\[0.5em]
$^{d}${\it LPHE-Modeling and Simulation, Faculty  of Sciences, Rabat, Morocco}\\[0.5em]
$^{e}${\it Centre of Physics and Mathematics,  Rabat, Morocco}\\[0.5em]

\vspace{3cm} {\bf Abstract}
\end{center}
\baselineskip=18pt
\medskip

The Hilbert-Schmidt distance between a mixed three-qubit state  and
its closest state is used to quantify the amount of pairwise quantum
correlations in a tripartite system. Analytical expressions of
geometric quantum discord are derived. A particular attention is
devoted to  two special classes of three-qubit $X$ states. They
include three-qubit states of ${\rm W}$, ${\rm GHZ}$ and Bell type.
We also discuss the monogamy property of geometric quantum discord
in some mixed three-qubit systems.

\vspace{1cm}

\newpage
\section{Introduction}

Quantum correlations in multipartite systems have been intensively
investigated  during the last two decades in the context of quantum
information science. This is mainly motivated by the fact that
quantum correlations constitute a key ressource for many quantum
information processing tasks (see for instance
\cite{NC-QIQC-2000,Horodecki,Guhne,Vedral-RMP-2002}). Also, the
understanding of the basic features of quantum correlations is
essential to provide a comprehensive way to distinguish the frontier
between quantum and classical physics. Nowadays, quantum
correlations have become an important tool in studying several
aspects in many-body systems such as quantum phase transition in
strongly correlated systems. A rigorous quantitative and qualitative
way to decide about the existence of quantum correlation, between
the compounds of a composite system, remains an open problem.
Various measures to quantify the degree of quantumness in
multipartite quantum systems have been discussed in the literature
from different perspectives and for several purposes (for a recent
review see \cite{Vedral-RMP-2012}).
  Among these several quantifiers of non-classicality,
concurrence and entanglement of formation \cite{Wootters98,Hil97}
have attracted considerable  attention.  But, recently it was
realized that entanglement of formation does not reveal  all non classical
aspects of quantum correlations. In this sense, quantum discord was
introduced to capture the essential of quantum correlations in
composite quantum systems. This measure, which goes beyond
entanglement of formation, is defined as the difference between the
total amount of nonclassical mutual information and classical
correlation present in a bipartite
 system \cite{Ollivier-PRL88-2001,Vedral-et-al}. The explicit expression of quantum discord requires an optimization procedure that
is in general a challenging task. To overcome this problem, a
geometric variant of quantum discord was proposed in
\cite{Dakic2010}. Geometric quantum discord was explicitly evaluated
between qubit-qubit as well as qubit-qudit systems (see \cite{Rau}
and references therein). In the literature, a particular attention
was devoted to quantum correlations in  the so-called two-qubit $X$ states [12-21, 23-27]
In the computational basis, these states have non-zero entries only
along the diagonal and anti-diagonal and look like the alphabet $X$.
Their algebraic structures \cite{Rau1}  simplify many analytical
calculation in deriving entanglement of formation \cite{Rau2} and
quantum discord \cite{Luo,Ali,rau3}. Interestingly,  algebraic
aspects of multi-qubit states have been generalized to describe $X$
states of quantum systems encompassing more than two qubits
\cite{van}. The generalized $X$ states cover a large class of
multi-qubit states including  ${\rm W}$ \cite{Werner,Zhang}, ${\rm
GHZ}$ \cite{Beth} and Dicke states \cite{Dixon}.\\

The study of genuine correlations in multipartite quantum systems is
complex from conceptual as well as computational point of view.
Various approaches, inspired by the results obtained of bipartite
systems, were discussed in the literature to tackle this issue. In
this paper, we extend the geometric measure of quantum discord for
two qubits, to tripartite systems comprising three qubits.  The
focus will be maintained strictly on  two special families of
three-qubit $X$ states for which the explicit expressions of quantum
discord are explicitly derived using the Hilbert-Schmidt norm. In
other hand, another important question in systems, comprising more
than two parts, concerns the distribution of quantum correlations
among the subsystems and it is constrained by the the so-called
monogamy relation. In fact, denoting by $Q$
a bipartite measure of quantum correlations in a tripartite system $1-2-3$, the sum
of quantum correlations $Q_{1\vert 2}$ (the shared correlation
between $1$ and $2$) and $Q_{1\vert 3}$ (the shared correlation
between $1$ and $3$) is always less or equal to  the correlation
$Q_{1\vert 23}$ shared between $1$ and the composite subsystem $23$.
The concept of monogamy was introduced by Coffman, Kundo and
Wootters in 2001 \cite{Coffman} in investigating the distribution of
entanglement in three qubit systems. The monogamy property was
analyzed for other measures of quantum correlations to understand
the distribution of correlations in multipartite systems and to
establish  the conditions limiting the shareability of quantum
correlations.  The
entanglement of formation \cite{Adesso2,Adesso3}, quantum discord
\cite{Giorgi,Prabhu,Sudha, Allegra,Ren} and its geometrized variant
\cite{Bruss,daoud2,daoud3} do not follows in general the  monogamy
property, contrarily to squared concurrence \cite{Coffman}.\\

This paper is organized as follows. In section 2, we introduce two
families of three-qubit $X$ states. The first one is given by three-qubit states where a subsystem comprising two qubits possesses parity
invariance. The second class corresponds to the situation where the
three qubits are all invariant under parity symmetry. In section 3, we
 derive the geometric measure of quantum discord. We also
give the explicit  forms of classical tripartite states presenting zero  discord. To investigate the
monogamy property in three-qubit $X$ states, we give the general
expression of geometric quantum discord in reduced states containing
two qubits after tracing-out the third  qubit in the global quantum
state. The explicit expressions of resulting pairwise quantum
discord are derived in section 4. To illustrate our calculations, we
consider some special instances of three qubit systems
for which geometric quantum correlations are given. In addition, we
discuss the distribution of geometric quantum discord to decide
about the monogamy property. Illustrations for some specify three-qubit mixed states
are given. Concluding
remarks close this paper.

\section{Three-qubit $X$ states}

$X$ states of two qubits have already found applications in many studies of
entanglement and discord [12-27]. As mentioned in the introduction, the interest in generalized $X$ states is motivated by the fact that they cover
many different states of interest in quantum information such as ${\rm W}$ and
${\rm GHZ}$ and Dicke states. The generalized $X$ states are of paramount importance
in investigating  quantum correlations for a collection of spin-$1/2$ particles possessing discrete symmetries like particle exchange
symmetry and/or parity invaraince. For instance, the reduced density matrices of multipartite Schr\"odinger cat states, which  are invariant under permutation symmetry, are  $X$ structured
 operators  (see for instance the reference \cite{Seddik}). Completely symmetric systems, including Dicke states, are relevant in many experimental situations such as spin squeezing which may have potential applications in atomic interferometers and high atomic clocks (see \cite{wang+sanders} and references therein) . Also, The multi-qubit $X$ states arise naturally in describing the dynamics tripartite quantum spin states interacting with a large environment \cite{J.Zhou}. This is of crucial importance in analyzing the decoherence effects induced by the environment in such systems.

In Fano-Bloch representation, a two  qubit  state writes as
\begin{equation}\label{fano-X}
\rho_{12} = \frac{1}{4} \sum_{\alpha,\beta = 0}^{4} T_{\alpha \beta} \sigma_{\alpha} \otimes \sigma_{\beta}
\end{equation}
where the Fano-Bloch parameters are given by $ T_{\alpha \beta} =
{\rm Tr} (\rho_{12} \sigma_{\alpha} \otimes \sigma_{\beta})$ and $\sigma_{\alpha}$ are the Pauli matrices. The
symmetry of two qubit systems is fully characterized  by the algebra
${\rm su(4)}$ spanned by the 4 $\times$ 4  Pauli matrices (see
\cite{Rau1,Rau2,rau3} and references therein). An interesting family
of two-qubit states which is relevant in several problems of quantum
optics and quantum information is the subset whose density matrices
resemble the letter $X$. They especially arise in physical systems
possessing parity symmetry such as Werner, Bell-diagonal and Dicke
states. The $X$ states are parameterized by seven real parameters
(three real parameters along the diagonal and two complex parameters
at off-diagonal positions). The underlying symmetry is characterized
by the sub-algebra ${\rm su(2)}\times {\rm u(1)} \times {\rm su(2)} \subset {\rm su(4)}$
spanned by seven linearly independent generators. Specifically, $X$
states can be written as
\begin{equation}
\rho_{12} =  \left(
\begin{array}{cccc}
\rho_{11} & 0 & 0 & \rho_{14} \\
0 & \rho_{22} & \rho_{23} & 0 \\
0 & \rho_{32} & \rho_{33} & 0 \\
\rho_{41} & 0 & 0 & \rho_{44}
\end{array}
\right).
\label{eqn1}
\end{equation}
in the computational basis for two qubits $(|00 \rangle, |01\rangle,
|10 \rangle, |11\rangle)$ or equivalently  $(|\uparrow \uparrow
\rangle, |\uparrow \downarrow \rangle, |\downarrow \uparrow \rangle,
|\downarrow \downarrow \rangle)$ in two spin basis. Clearly, the
states of the form (\ref{eqn1}) commute with the operator
$\sigma_3\otimes\sigma_3$ reflecting the invariance under parity
transformation. The tools developed for two  qubit systems are of
paramount importance for three or more qubits. The  $X$ states for
multi-qubit systems and their underlying symmetries were discussed
in \cite{Rau,Rau1,Rau2,rau3}. In this paper we shall mainly focus on
three-qubits $X$ states. We consider a tripartite system $1-2-3$
with each party holding a qubit. The state shared between three
parties $1$, $2$ and $3$ is given by the unit trace operator
$\rho_{123}$ acting on the tensor-product Hilbert space ${\cal H}_1
\otimes {\cal H}_2 \otimes {\cal H}_3 $ where each  single Hilbert
space is two-dimensional spanned by the vectors $\vert 0 \rangle$
and $\vert 1 \rangle$. The  three qubit system lives in a
$2^3$-dimensional Hilbert space. As mentioned in the introduction,
two types  of $X$ states are studied  in this work.  The first type
concerns the states commuting with the operator
$\sigma_3\otimes\sigma_3 \otimes \sigma_0$ and the second class
corresponds to density matrices that commute with the operators
$\sigma_3\otimes\sigma_3 \otimes \sigma_3$.

\subsection{ Three-qubit $X$ states: first class}

The first family of three-qubit states, that we introduce in this
section, corresponds to density matrices  commuting $\sigma_3
\otimes \sigma_3 \otimes \sigma_0$. The states of the subsystem $1-2$  of the tripartite system $1-2-3$ are invariant under
parity transformation. It is simply verified that, in the usual
$2^3$-dimensional computational basis, the general form of such states is

\begin{equation}\label{3X}
\rho_{123}=\left(%
\begin{array}{cccccccc}
\rho_{11} & 0 & 0 & \rho_{14} & \rho_{15} & 0 & 0 & \rho_{18} \\
  0 & \rho_{22} & \rho_{23} & 0 & 0  & \rho_{26} & \rho_{27} & 0 \\
  0 & \rho_{32} & \rho_{33} & 0 & 0 & \rho_{36} & \rho_{37} & 0 \\
  \rho_{41} & 0 & 0 & \rho_{44} &  \rho_{45} & 0 & 0 &  \rho_{48} \\
  \rho_{51} & 0 & 0 & \rho_{54} & \rho_{55} & 0 & 0 & \rho_{58} \\
 0 &  \rho_{62}&  \rho_{63} & 0 & 0 & \rho_{66} & \rho_{67} & 0 \\
  0 & \rho_{72} & \rho_{73}& 0 & 0 & \rho_{76} & \rho_{77} & 0 \\
 \rho_{81} & 0 & 0 & \rho_{84} & \rho_{85} & 0 & 0 & \rho_{88} \\
\end{array}%
\right).
\end{equation}
In the Fano-Bloch representation,  the three-qubit state (\ref{3X})
takes the following form
\begin{equation}\label{123rho}
\rho_{123}=\frac{1}{8} \sum_{\alpha \beta \gamma }{\cal R}_{\alpha
\beta \gamma}~ \sigma_\alpha\otimes\sigma_\beta\otimes \sigma_\gamma
\end{equation}
where $\alpha$,$\beta$ and $\gamma$ take the values =$0,1,2,3$ and the the correlation matrix
elements ${\cal R}_{\alpha\beta\gamma}$ are
$$ {\cal R}_{\alpha\beta\gamma} = {\rm Tr} \big( \rho_{123}  (\sigma_{\alpha} \otimes \sigma_{\beta} \otimes \sigma_{\gamma})\big) $$
with ${\cal R}_{000} = 1$ (${\rm Tr}(\rho_{123})=1$). The operators
$\sigma_{\alpha}$ stands for Pauli basis with $\sigma_0$ is the
identity. The parity invariance reduces the number of the non
vanishing  correlation matrix elements ${\cal
R}_{\alpha\beta\gamma}$ in equation (\ref{123rho}). Indeed, it is
easy to verify that the non vanishing ones are those corresponding
to $(\alpha, \beta, \gamma)$ belonging to the following set of
triplets
\begin{equation}
(000), (001), (002), (003), (030), (031), (032), (033)$$
 $$ (110), (111), (112), (113), (120), (121), (122), (123)$$
 $$ (210), (211), (212), (213), (220), (221), (222), (223)$$
 $$(300) , (301), (302), (303), (330), (331), (332), (333).
\end{equation}
Accordingly,  the  state (\ref{123rho}) expand as
\begin{equation}\label{3Xfano}
\rho_{123}=\frac{1}{8}\bigg[{\cal R}_{000} \sigma_0\otimes
\sigma_0\otimes \sigma_0 + \sum_i\big( {\cal
R}_{i00}~\sigma_i\otimes \sigma_0\otimes \sigma_0+ {\cal R}_{0i0} ~
\sigma_0\otimes\sigma_i\otimes \sigma_0+ {\cal R}_{00i}
~\sigma_0\otimes \sigma_0\otimes\sigma_i\big)$$ $$+\sum_{ij}
\big({\cal R}_{ij0}~\sigma_i\otimes \sigma_j\otimes \sigma_0 + {\cal
R}_{i0j} ~\sigma_i \otimes \sigma_0\otimes\sigma_j+{\cal R}_{0ij}~
\sigma_0\otimes\sigma_i\otimes \sigma_j \big)+\sum_{ijk}{\cal
R}_{ijk}~ \sigma_i\otimes\sigma_j\otimes \sigma_k \bigg],
\end{equation}
in terms of 32 operators which span the subalgebra $ {\rm su(2)} \otimes
{\rm u(1)} \otimes {\rm su(2)} \otimes {\rm u(1)} \otimes {\rm su(2)} \otimes {\rm u(1)} \otimes
{\rm su(2)} $ of the full symmetry algebra ${\rm su(8)}$ characterizing an
arbitrary three-qubit system \cite{Rau1,Rau2,rau3}. The explicit relation between the non
vanishing Fano-Bloch parameters ${\cal R}_{\alpha\beta\gamma}$ and
the matrix elements of
 $\rho_{123}$ will be given here after. It is interesting to note
 that density matrix (\ref{3X}) encompasses  four two qubit $X$
 states (four sub-blocks, each one is $X$ shaped). In fact, the matrix (\ref{3X}) can
be written as
\begin{equation}\label{3X-fano-1}
\rho_{123}= \sum_{i,j = 0,1} \rho^{ij} \otimes \vert i \rangle
\langle j \vert
\end{equation}
where the vectors $\vert i \rangle$ and $\vert j \rangle $ are
related to the qubit $3$. From equation (\ref{3X}), the density
matrices $\rho^{ij}$ appearing in (\ref{3X-fano-1})  write as
\begin{eqnarray}
\rho^{ij} = \left(
\begin{array}{cccc}
\rho_{1+4i~ 1+4j}
& 0 & 0 & \rho_{1+4i~ 4+4j} \\
0 & \rho_{2+4i~ 2+4j} & \rho_{2+4i~3+4j} & 0 \\
0 & \rho_{3+4i~ 2+4j} & \rho_{3+4i~3+4j}
& 0 \\
\rho_{4+4i~1+4j} & 0 & 0 & \rho_{4+4i~4+4j}
\end{array}
\right) \,, \label{2X}
\end{eqnarray}
 in the computational basis spanned by two-qubit product states of
$1$ and $2$ $\{  |0 \rangle_1 \otimes |0\rangle_2, |0\rangle_1
\otimes |1\rangle_2 , |1\rangle_1 \otimes |0\rangle_2, |1\rangle_1
\otimes |1\rangle_2\} $.  The Fano-Bloch representations of the two
qubit $X$ states $\rho^{ij}$ (\ref{2X}) are
\begin{equation}\label{2X-fano}
\rho^{ij}=\frac{1}{4}  \sum_{\alpha \beta} R^{ij}_{\alpha \beta} \sigma_{\alpha}\otimes \sigma_{\beta}
\end{equation}
where $\alpha, \beta = 0,1,2,3$ and the Fano-Bloch parameters
$R^{ij}_{\alpha \beta}$ defined by
$$R^{ij}_{\alpha \beta} = {\rm Tr}(\rho^{ij}~\sigma_{\alpha}\otimes \sigma_{\beta}),$$
are given by
\begin{equation}\label{Rij}
R^{ij}_{00}= 1$$
$$ R^{ij}_{30}=\rho_{1+4i~ 1+4j}+\rho_{2+4i~ 2+4j}-\rho_{3+4i~3+4j}-\rho_{4+4i~4+4j}$$
$$R^{ij}_{03}=\rho_{1+4i~ 1+4j}-\rho_{2+4i~ 2+4j}+\rho_{3+4i~3+4j}-\rho_{4+4i~4+4j}$$
$$R^{ij}_{11}=\rho_{1+4i~ 4+4j}+\rho_{4+4i~1+4j}+\rho_{2+4i~3+4j}+\rho_{3+4i~ 2+4j}$$
$$R^{ij}_{12}=i(\rho_{1+4i~ 4+4j}-\rho_{4+4i~1+4j}-\rho_{2+4i~3+4j}+\rho_{3+4i~ 2+4j})$$
$$R^{ij}_{21}=i(\rho_{1+4i~ 4+4j}-\rho_{4+4i~1+4j}+\rho_{2+4i~3+4j}-\rho_{3+4i~ 2+4j})$$
$$R^{ij}_{22}=\rho_{2+4i~3+4j}+\rho_{3+4i~ 2+4j}-\rho_{1+4i~ 4+4j}-\rho_{4+4i~1+4j}$$
$$R^{ij}_{33}=\rho_{1+4i~ 1+4j}-\rho_{2+4i~ 2+4j}-\rho_{3+4i~3+4j} +
\rho_{4+4i~4+4j}).
\end{equation}
By inserting  the Fano-Bloch representations (\ref{2X-fano}) into
the expression (\ref{3X-fano-1}), the tripartite correlations
elements ${\cal R}_{\alpha\beta\gamma}$ can be written in terms of
the bipartite correlation parameters $R^{ij}_{\alpha \beta}$.
Indeed, equation (\ref{3X-fano-1}) can be rewritten as
\begin{equation}\label{3X-fano1}
\rho_{123}=\frac{1}{2} \Bigg[ (\rho^{00} + \rho^{11} )\otimes
\sigma_0  +   (\rho^{00} - \rho^{11} )\otimes \sigma_3 +
 (\rho^{01} + \rho^{10} )\otimes \sigma_1  + i (\rho^{01} - \rho^{10} )\otimes \sigma_2 \Bigg]
\end{equation}
and similarly, we rewrite (\ref{3Xfano}) as
\begin{equation}\label{3Xfano11}
\rho_{123} =  \frac{1}{8} \sum_{\alpha\beta} \Bigg[{\cal
R}_{\alpha\beta0} ~\sigma_{\alpha} \otimes \sigma_{\beta} \otimes
\sigma_{0}+ {\cal R}_{\alpha\beta1} ~\sigma_{\alpha} \otimes
\sigma_{\beta} \otimes \sigma_{1}+ {\cal R}_{\alpha\beta2}
~\sigma_{\alpha} \otimes \sigma_{\beta} \otimes \sigma_{2}+ {\cal
R}_{\alpha\beta3} ~\sigma_{\alpha} \otimes \sigma_{\beta} \otimes
\sigma_{3}\Bigg].
\end{equation}
By Replacing  the expressions (\ref{2X-fano}) and (\ref{Rij})  in
(\ref{3X-fano1}), and identifying with the equation
(\ref{3Xfano11}), one gets
\begin{equation}\label{relation3ret2r}
{\cal R}_{\alpha\beta0}=    R^{++}_{\alpha\beta} = R^{00}_{\alpha\beta} + R^{11}_{\alpha\beta}  $$
$$ {\cal R}_{\alpha\beta3}= R^{--}_{\alpha\beta} = R^{00}_{\alpha\beta} - R^{11}_{\alpha\beta}$$
$$ {\cal R}_{\alpha\beta1} = R^{+-}_{\alpha\beta} = R^{01}_{\alpha\beta} + R^{10}_{\alpha\beta}$$
$$ {\cal R}_{\alpha\beta2} = R^{-+}_{\alpha\beta} = iR^{01}_{\alpha\beta} - iR^{10}_{\alpha\beta}.
\end{equation}
where  the pairs $(\alpha \beta)$ belong to the set $ \{ (00), (03),
(30), (12) , (21) , (11), (22), (33) \}.$ The relations
(\ref{relation3ret2r}) specify  completely the tripartite
correlation  tensor ${\cal R}_{\alpha\beta\gamma}$  in terms  of the
Fano-Bloch parameters $R^{ij}_{\alpha \beta}$ encoding the
correlations in the two qubit subsystem $1-2$ (\ref{2X-fano}). As we
shall discuss, these recursive relations play a central role
in deriving the geometric measure of quantum discord.

\subsection{Three-qubit $X$ states: second class}

Now we consider  three-qubit states, denoted by $\sigma_{123}$,  possessing the
symmetry invariance under the parity transformation
$\mathbb{Z}_2\otimes \mathbb{Z}_2\otimes \mathbb{Z}_2$. As they commute
 with the parity operator
$\sigma_3\otimes\sigma_3\otimes\sigma_3$,  they write
 \begin{equation}\label{3X-class2}
\sigma_{123}=\left(%
\begin{array}{cccccccc}
\sigma_{11} & 0 & 0 & \sigma_{14} & 0 & \sigma_{16} & \sigma_{17} & 0 \\
  0 & \sigma_{22} & \sigma_{23} & 0 & \sigma_{25} & 0 & 0 & \sigma_{28} \\
  0 & \sigma_{32} & \sigma_{33} & 0 & \sigma_{35} & 0 & 0 & \sigma_{38} \\
  \sigma_{41} & 0 & 0 & \sigma_{44} & 0 & \sigma_{46} & \sigma_{47} & 0 \\
  0 & \sigma_{52} & \sigma_{53} & 0 & \sigma_{55} & 0 & 0 & \sigma_{58} \\
  \sigma_{61} & 0 & 0 & \sigma_{64} & 0 & \sigma_{66} & \sigma_{67} & 0 \\
  \sigma_{71} & 0 & 0 & \sigma_{74} & 0 & \sigma_{76} & \sigma_{77} & 0 \\
  0 & \sigma_{82} & \sigma_{83} & 0 & \sigma_{85} & 0 & 0 & \sigma_{88} \\
\end{array}%
\right)
\end{equation}
 in the standard computational basis. The density matrix $\sigma_{123}$ is built of four blocks. The diagonal blocks appear as $X$
alphabet with non-zero
density matrix elements only along the diagonal and anti-diagonal contrarily to  the
two off diagonal blocks which have
vanishing  elements  along the diagonal and anti-diagonal. This gives  another family  of  extended three-qubit $X$ state (see \cite{Rau1,Rau2,rau3}
where such states were originally termed $X$ states). The underlying
symmetry is $ {\rm su(2)} \otimes {\rm u(1)} \otimes {\rm su(2)} \otimes {\rm u(1)} \otimes
{\rm su(2)} \otimes {\rm u(1)} \otimes {\rm su(2)}$. In the Fano-Bloch representation,
the matrix density (\ref{3X-class2}) expands as
\begin{equation}\label{3X-fano-class2}
\sigma_{123}=\frac{1}{8}\bigg[{\cal T}_{000} \sigma_0\otimes
\sigma_0\otimes \sigma_0 + \sum_i\big( {\cal
T}_{i00}~\sigma_i\otimes \sigma_0\otimes \sigma_0+ {\cal T}_{0i0} ~
\sigma_0\otimes\sigma_i\otimes \sigma_0+ {\cal T}_{00i}
~\sigma_0\otimes \sigma_0\otimes\sigma_i\big)$$ $$+\sum_{ij}
\big({\cal T}_{ij0}~\sigma_i\otimes \sigma_j\otimes \sigma_0 + {\cal
T}_{i0j} ~\sigma_i \otimes \sigma_0\otimes\sigma_j+{\cal T}_{0ij}~
\sigma_0\otimes\sigma_i\otimes \sigma_j \big)+\sum_{ijk}{\cal
T}_{ijk}~ \sigma_i\otimes\sigma_j\otimes \sigma_k \bigg]
\end{equation}
where the the matrix correlation elements are
$$ {\cal T}_{\alpha\beta\gamma} = {\rm Tr} \big( \sigma_{123}  (\sigma_{\alpha} \otimes \sigma_{\beta} \otimes \sigma_{\gamma})\big) $$
with ${\cal T}_{000} = 1$ (${\rm Tr}(\sigma_{123})=1$). The non
vanishing correlation elements ${\cal T}_{\alpha\beta\gamma}$
occurring in (\ref{3X-fano-class2}) are those with a triplet
$(\alpha\beta\gamma)$ in the following list
\begin{equation}\label{set2}
(000), (003), (011), (012), (021), (022), (030), (033)$$
 $$ (101), (102), (110), (113), (120), (123), (131), (132)$$
 $$ (201), (202), (210), (213), (220), (223), (231), (232)$$
 $$(300) , (303), (311), (312), (321), (322), (330), (333).
\end{equation}
Analogously to the previous class of three-qubit states, we write the density
matrix  (\ref{3X-class2}) as
\begin{equation}\label{3X-fano1-calss2}
\sigma_{123}= \sum_{i,j = 0,1} \sigma^{ij} \otimes \vert i \rangle
\langle j \vert
\end{equation}
where $\vert i \rangle$ and  $\vert j \rangle $ are eigenvectors
associated with the third qubit.  In equation
(\ref{3X-fano1-calss2}), the matrices $\sigma^{ii}$ (with $i
= 0,1$) write, in the computational basis spanned by two-qubit
product states of the subsystems $1$ and $2$ $\{ |0 \rangle_1
\otimes |0\rangle_2, |0\rangle_1 \otimes |1\rangle_2 , |1\rangle_1
\otimes |0\rangle_2, |1\rangle_1 \otimes |1\rangle_2\} $, as
\begin{eqnarray}
\sigma^{ii} = \left(
\begin{array}{cccc}
\sigma_{1+4i~ 1+4i}
& 0 & 0 & \sigma_{1+4i~ 4+4i} \\
0 & \sigma_{2+4i~ 2+4i} & \sigma_{2+4i~3+4i} & 0 \\
0 & \sigma_{3+4i~ 2+4i} & \sigma_{3+4i~3+4i}
& 0 \\
\sigma_{4+4i~1+4i} & 0 & 0 & \sigma_{4+4i~4+4i}
\end{array}
\right) \,. \label{2X-class2}
\end{eqnarray}
For $(i = 0 , j = 1)$ and $(i = 1 , j = 0)$, we have
\begin{eqnarray}
\sigma^{ij} = \left(
\begin{array}{cccc}
0
& \sigma_{1+4i~ 2+4j} & \sigma_{1+4i~ 3+4j} & 0 \\
\sigma_{2+4i~ 1+4j} & 0 & 0  & \sigma_{2+4i~ 4+4j} \\
\sigma_{3+4i~ 1+4j} & 0 & 0
& \sigma_{3+4i~ 4+4j}\\
0 & \sigma_{4+4i~ 2+4j} & \sigma_{4+4i~ 3+4j} & 0
\end{array}
\right) \,. \label{2X-class2}
\end{eqnarray}
The Fano-Bloch representations of the matrices $\sigma^{ii}$ are
\begin{equation}\label{2X-fano-class2-1}
\sigma^{ii}=\frac{1}{4}  \sum_{\alpha \beta} T^{ii}_{\alpha \beta} ~ \sigma_{\alpha}\otimes \sigma_{\beta}
\end{equation}
where $\alpha, \beta = 0,1,2,3$ and the
vanishing correlation parameters  $T^{ij}_{\alpha \beta}$ are given
by
\begin{equation}\label{Tii}
T^{ii}_{00}= 1$$
$$ T^{ii}_{30}=\sigma_{1+4i~ 1+4i}+\sigma_{2+4i~ 2+4i}-\sigma_{3+4i~3+4i}-\sigma_{4+4i~4+4i}$$
$$T^{ii}_{03}=\sigma_{1+4i~ 1+4i}-\sigma_{2+4i~ 2+4i}+\sigma_{3+4i~3+4i}-\sigma_{4+4i~4+4i}$$
$$T^{ii}_{11}=\sigma_{1+4i~ 4+4i}+\sigma_{4+4i~1+4i}+\sigma_{2+4i~3+4i}+\sigma_{3+4i~ 2+4i}$$
$$T^{ii}_{12}=i(\sigma_{1+4i~ 4+4i}-\sigma_{4+4i~1+4i}-\sigma_{2+4i~3+4i}+\sigma_{3+4i~ 2+4i})$$
$$T^{ii}_{21}=i(\sigma_{1+4i~ 4+4i}-\sigma_{4+4i~1+4i}+\sigma_{2+4i~3+4i}-\sigma_{3+4i~ 2+4i})$$
$$T^{ii}_{22}=\sigma_{2+4i~3+4i}+\sigma_{3+4i~ 2+4i}-\sigma_{1+4i~ 4+4i}-\sigma_{4+4i~1+4i}$$
$$T^{ii}_{33}=\sigma_{1+4i~ 1+4i}-\sigma_{2+4i~ 2+4i}-\sigma_{3+4i~3+4i} +
\sigma_{4+4i~4+4i}).
\end{equation}
Similarly, for the two-qubit matrices $\sigma^{ij}$ (\ref{2X-class2}), the
corresponding Fano-Bloch representations are
\begin{equation}\label{2X-fano-class2-2}
\sigma^{ij}=\frac{1}{4}  \sum_{\alpha \beta} T^{ij}_{\alpha \beta} ~ \sigma_{\alpha}\otimes \sigma_{\beta} \qquad i\neq j
\end{equation}
where the non zero matrix elements $T^{ij}_{\alpha \beta}$ are given
by
\begin{equation}\label{Tij}
T^{ij}_{01}=\sigma_{2+4i~ 1+4j}+\sigma_{1+4i~ 2+4j}+\sigma_{4+4i~ 3+4j}+\sigma_{3+4i~ 4+4j}$$
$$T^{ij}_{02}=i(-\sigma_{2+4i~ 1+4j}+\sigma_{1+4i~ 2+4j}-\sigma_{4+4i~ 3+4j}+\sigma_{3+4i~ 4+4j})$$
$$T^{ij}_{10}=\sigma_{1+4i~ 3+4j}+\sigma_{3+4i~ 1+4j}+\sigma_{2+4i~ 4+4j}+\sigma_{4+4i~ 2+4j}$$
$$T^{ij}_{13}=\sigma_{1+4i~ 3+4j}+\sigma_{3+4i~ 1+4j}-\sigma_{2+4i~ 4+4j}-\sigma_{4+4i~ 2+4j}$$
$$T^{ij}_{20}=i(\sigma_{1+4i~ 3+4j}-\sigma_{3+4i~ 1+4j}+\sigma_{2+4i~ 4+4j}-\sigma_{4+4i~ 2+4j})$$
$$T^{ij}_{23}=i(\sigma_{1+4i~ 3+4j}-\sigma_{3+4i~ 1+4j}-\sigma_{2+4i~ 4+4j}+\sigma_{4+4i~ 2+4j})$$
$$T^{ij}_{31}=\sigma_{1+4i~ 2+4j}+\sigma_{2+4i~ 1+4j}-\sigma_{3+4i~ 4+4j}-\sigma_{4+4i~ 3+4j})$$
$$T^{ij}_{32}=i(\sigma_{1+4i~ 2+4j}-\sigma_{2+4i~ 1+4j}-\sigma_{3+4i~ 4+4j}+\sigma_{4+4i~
3+4j}).
\end{equation}
Using (\ref{3X-fano1-calss2}), one obtains
\begin{equation}\label{3X-fano2}
\sigma_{123}=\frac{1}{2} \Bigg[ (\sigma^{00} + \sigma^{11} )\otimes
\sigma_0  +   (\sigma^{00} - \sigma^{11} )\otimes \sigma_3 +
 (\sigma^{01} + \sigma^{10} )\otimes \sigma_1  + i (\sigma^{01} - \sigma^{10} )\otimes \sigma_2
 \Bigg].
\end{equation}
The three-qubit state (\ref{3X-fano-class2}) rewrites also as
\begin{equation}
\sigma_{123} = \frac{1}{8} \sum_{\alpha\beta} \Bigg[{\cal
T}_{\alpha\beta0} ~\sigma_{\alpha} \otimes \sigma_{\beta} \otimes
\sigma_{0}+ {\cal T}_{\alpha\beta1} ~\sigma_{\alpha} \otimes
\sigma_{\beta} \otimes \sigma_{1}+ {\cal T}_{\alpha\beta2}
~\sigma_{\alpha} \otimes \sigma_{\beta} \otimes \sigma_{2}+ {\cal
T}_{\alpha\beta3} ~\sigma_{\alpha} \otimes \sigma_{\beta} \otimes
\sigma_{3}\Bigg]
\end{equation}
Inserting  (\ref{2X-fano-class2-1}) and  (\ref{2X-fano-class2-2})
in the equation (\ref{3X-fano2}),
one verifies the following relations
\begin{equation}\label{relation3ret2r-class2}
{\cal T}_{\alpha\beta0}=    T^{++}_{\alpha\beta} = T^{00}_{\alpha\beta} + T^{11}_{\alpha\beta}  $$
$$ {\cal T}_{\alpha\beta3}= T^{--}_{\alpha\beta} = T^{00}_{\alpha\beta} - T^{11}_{\alpha\beta}
\end{equation}
where  $(\alpha \beta)$ belongs to the set $\{ (00), (03), (30),
(12) , (21) , (11), (22), (33) \}$ and
\begin{equation}\label{relation3ret2r-class22}
{\cal T}_{\alpha\beta1} = T^{+-}_{\alpha\beta} = T^{01}_{\alpha\beta} + T^{10}_{\alpha\beta}$$
$$ {\cal T}_{\alpha\beta2} = T^{-+}_{\alpha\beta} = iT^{01}_{\alpha\beta} - iT^{10}_{\alpha\beta}
\end{equation}
where  $(\alpha \beta)$ are in the set $ \{(01), (02), (10),
(20) , (13) , (23), (31), (32)\}$, so that the total number of non
vanishing correlation matrix ${\cal T}_{\alpha\beta\gamma}$ elements
is 32 to be compared with (\ref{set2}). The relations (\ref{relation3ret2r-class2}) and (\ref{relation3ret2r-class22}) reflect
that the tensor element ${\cal T}_{\alpha\beta\gamma}$  can be explicitly expressed in terms of two-qubit correlations factors.

\section{Geometric measure of quantum discord}
A bipartite quantum system exhibits quantum correlation if its two subsystems contain more information than taken separately. This concept is captured by
the mutual information  $I(A:B)=H(A)+H(B) - H(A,B)$
where $A$ and $B$ are random variables. In classical information theory $H(.)$ stands
 for the Shannon entropy $H(p)= - \sum_i p_i\log p_i$ where $p=(p_1, p_2, \cdots)$ is the probability distribution. For a quantum density matrix $\rho$,  $H(.)$  denotes the von Neumann entropy  $H(\rho)= -{\rm Tr}\rho\log \rho$. In the classical case, an equivalent expression for the mutual information is given by   $I(A:B)=H(A)- H(A|B)$ where $H(A|B)$ is the Shannon entropy of
$A$ conditioned on the measurement outcome of $B$. In the quantum case, the two expressions are different and the difference defines the so-called quantum discord \cite{Ollivier-PRL88-2001,Vedral-et-al}. The von Neumann entropy-type quantum discord involves complicated optimization procedures \cite{Ali}. In the literature there are few examples for which closed analytical expressions for quantum discord were obtained (see the review  \cite{Vedral-RMP-2012}). Alternatively, distance-type quantifiers of quantum discord have been considered. This is essentially motivated by their presumably simple evaluation in comparison with the original quantum discord definition. Several distances are possible (trace distance, Bures distance, ...)
with their own advantages and drawbacks. In this paper we shall especially consider the
geometric discord variant based on Hilbert-Schmidt norm \cite{Dakic2010}. Thus, given a tripartite system $1-2-3$, we shall consider the bipartite splitting $1\vert 23$. The pairwise quantum correlation between the subsystems $(1)$ and $(23)$ in three-qubit $X$ states of type (\ref{3X}) or (\ref{3X-class2}) is determined
in complete analogy with two qubit $X$ state. It is defined as the distance from the set of classically correlated states using Hilbert-Schmidt trace. In this respect, the explicit
form of states of type (\ref{3X}) or (\ref{3X-class2}) presenting vanishing quantum correlation can be  derived by optimizing the Hilbert-Schmidt norm by means of which
quantunmness is quantified.  This issue constitutes the main of this section.

\subsection{Closest classical states to two qubit $X$ states }
To begin, we shall present the procedure leading to  the closest classically correlated state to the two-qubit $X$ state ({\ref{eqn1}}).
 The Fano-Bloch representation (\ref{fano-X})
reads
\begin{equation}\label{12X}
\rho_{12} = \frac{1}{4} \bigg[  \sigma_{0}\otimes \sigma_{0} +   T_{03}
\sigma_{0}\otimes \sigma_{3} +  T_{30}
\sigma_{3}\otimes \sigma_{0} + \sum_{kl} T_{kl}
\sigma_{k}\otimes \sigma_{l} \bigg]
\end{equation}
where the correlation matrix elements are obtainable from (\ref{Rij})
modulo some obvious substitutions.
The geometric measure of quantum discord is defined as the distance
the state $\rho_{12}$ and its closest classical-quantum state
presenting zero discord \cite{Dakic2010}
\begin{equation}\label{definition-discord}
D_{\rm g}(\rho_{12}) = \min_{\chi_{12}} \vert\vert \rho_{12} -
\chi_{12} \vert\vert^2
\end{equation}
where the Hilbert-Schmidt norm is defined by $ \vert\vert X
\vert\vert^2 = {\rm Tr} (X^{\dagger}X)$ and the minimization is
taken over the set of all classical states. When the measurement is
performed on the  qubit 1, the classical states
write
\begin{equation}
\chi_{12} = p_1 \vert \psi_1 \rangle \langle \psi_1 \vert \otimes \rho_1^{2} + p_2 \vert \psi_2 \rangle \langle \psi_2 \vert \otimes \rho_2^{2}
\end{equation}
where $\{ \vert \psi_1 \rangle , \vert \psi_2 \rangle \}$ is an
orthonormal basis related to the qubit $1$,  $p_i$ $(i=1,2)$ stands
for probability distribution and $\rho_i^{2}$ $(i=1,2)$ is the
marginal density of the qubit 2. The classically correlated states
$\chi_{12} $ can also be written as
\begin{equation}\label{xi12}
\chi_{12} = \frac{1}{4}\Bigg[ \sigma_{0}\otimes \sigma_{0} +
\sum_{i=1}^3 te_i ~~\sigma_i \otimes \sigma_{0} + \sum_{i=1}^3
(s_+)_{i} ~\sigma_0 \otimes \sigma_{i} + \sum_{i,j=1}^3  e_i
(s_-)_{j} ~\sigma_i \otimes \sigma_{j}\Bigg]
\end{equation}
where
$$ t = p_1-p_2, \qquad e_i = \langle \psi_1 \vert \sigma_i \vert \psi_1 \rangle,  \qquad
(s_{\pm})_{j} = {\rm Tr}\big((p_1\rho_1^{2} \pm p_2\rho_2^{2})
\sigma_{j} \big).$$
It follows that the distance between the density matrix $\rho_{12}$
and the classical state $\chi_{12}$, as measured by Hilbert-Schmidt
norm, is then given by
\begin{equation}\label{HS12-1}
|| \rho_{12} - \chi_{12}||^2 = \frac{1}{4} \bigg[  (t^2 -2te_3 T_{30} + T^2_{30}) +  \sum_{i=1}^{3}
(T_{0i} -  (s_+)_{i})^2 + \sum_{i,j=1}^3   ( T_{ij} -  e_i(s_-)_{j})^2  \bigg]
\end{equation}
The  minimization of the distance (\ref{HS12-1}),  with respect to the parameters $t$, $(s_+)_{i}$ and
$(s_-)_{i}$, gives
\begin{equation}\label{minHS12}
t= e_3T_{30} $$
$$ (s_+)_{1}= 0 \quad (s_+)_{2} = 0 \quad  (s_+)_{3} = T_{03} $$
$$ (s_-)_{i} = \sum_{j=1}^3 e_j T_{ji}.
\end{equation}
Inserting  these solutions in (\ref{HS12-1}), one has
\begin{equation}\label{HS12}
|| \rho_{12} - \chi_{12}||^2 = \frac{1}{4} \bigg[  {\rm Tr}K - \vec{e}^t K \vec{e} \bigg]
\end{equation}
where the matrix $K$ is defined by
\begin{equation}\label{matrixK12}
K = xx^{\dagger} + TT^{\dagger}
\end{equation}
with
$$ x^{\dagger}  = (0,0,T_{30})\qquad T = \left(%
\begin{array}{ccc}
T_{11} &  T_{12} & 0 \\
T_{21}   & T_{22} & 0 \\
0 & 0 &T_{33}  \\
\end{array}%
\right).$$
From equation (\ref{HS12}), one see that the minimal value of
Hilbert-Schmidt distance (\ref{HS12}) is reached for the largest
eigenvalue of the matrix $K$. We denote by $\lambda_1$, $\lambda_2$ and $\lambda_3$ the eigenvalues
of the matrix $K$ (\ref{matrixK12}) corresponding to the $X$ state
(\ref{eqn1})  or equivalently (\ref{12X}).  They
 are given by
\begin{equation}\label{vp:lambda12}
\lambda_1 = 4(\vert \rho_{14}\vert +\vert \rho_{23}\vert )^2,\qquad
\lambda_2 = 4(\vert \rho_{14} \vert - \vert \rho_{23}\vert )^2,
\qquad \lambda_3= 2[(\rho_{11}- \rho_{33})^2 + (\rho_{22}-
\rho_{44})^2].
\end{equation}
To get the minimal value of the Hilbert-Schmidt distance (\ref{HS12}) and subsequently the amount
of geometric quantum discord, one compares
$\lambda_1$, $\lambda_2$ and $\lambda_3$. As $\lambda_1$ is always greater than  $\lambda_2$,
the largest eigenvalue  $\lambda_{\rm max}$ is $\lambda_1$ or
$\lambda_3$. It follows that the geometric discord is given by
\begin{equation}
D_g(\rho_{12}) = \frac{1}{4}~ {\rm min}\{ \lambda_1 + \lambda_2 ,
\lambda_2 + \lambda_3\}.\label{eq:GMQD_new}
\end{equation}
To write down the  explicit  expressions of the closest classical
state $\chi_{12}$ to $\rho_{12}$, one has to determine the
eigenvector $\vec{e}_{\rm max}$ associated with the largest
eigenvalue $\lambda_{\rm max}$. In this respect, two cases ( $\lambda_{\rm max} =
\lambda_1$ and $\lambda_{\rm max} = \lambda_3$) are separately
discussed. We begin by density matrices $\rho_{12}$ (\ref{eqn1}) whose entries satisfy
the condition $\lambda_{\rm max} = \lambda_3$. The associated
eigenvector  is given by $\vec{e}_3 = (0,0,1)$. Replacing in the set
of  constraints (\ref{minHS12}), one has
\begin{equation}
\chi^3_{12} = \frac{1}{4}\Bigg[ \sigma_{0}\otimes \sigma_{0} +
T_{30} ~~\sigma_3 \otimes \sigma_{0} + T_{03}~\sigma_0 \otimes
\sigma_{3} + T_{33}~\sigma_3 \otimes \sigma_{3}\Bigg]
\end{equation}
In the second situation, the eigenvector corresponding to
$\lambda_1$ is given by $\vec{e}_1 = (\cos \frac{\phi}{2}, -
\sin\frac{\phi}{2} , 0)$ where $e^{i\phi} =
\frac{\rho_{14}\rho_{23}}{|\rho_{14}||\rho_{23}|} $. Reporting the
components of $\vec{e}_1$ in (\ref{minHS12}), one gets the closest
classical state
\begin{equation}
\chi^1_{12} = \frac{1}{4}\Bigg[ \sigma_{0}\otimes \sigma_{0} +
T_{30} ~~\sigma_3 \otimes \sigma_{0} + \sum_{i=1}^{2} \sum_{j=1}^{2}
\tilde{T}_{ij}~\sigma_i \otimes \sigma_{j} \Bigg]
\end{equation}
where
$$ \tilde{T}_{11} = \cos \frac{\phi}{2} (\cos \frac{\phi}{2} T_{11} - \sin \frac{\phi}{2} T_{21} )
\qquad  \tilde{T}_{12} =  \cos \frac{\phi}{2} (\cos \frac{\phi}{2}
T_{12} - \sin \frac{\phi}{2} T_{22} )$$
$$  \tilde{T}_{21} =  -\sin \frac{\phi}{2} (\cos \frac{\phi}{2} T_{11} - \sin \frac{\phi}{2} T_{21} )
\qquad \tilde{T}_{22} =  -\sin\frac{\phi}{2} (\cos \frac{\phi}{2}
T_{12} - \sin \frac{\phi}{2} T_{22} ).$$
As we already mentioned, the geometric quantifiers of quantum correlations in bipartite systems
can be extended to embrace  three-qubit $X$ states of type (\ref{3X}) or (\ref{3X-class2}).

\subsection{ Closest classical states to three-qubits $X$ states}
Along similar lines of reasoning, we determine first the closest classical
states to generalized $X$ states of the form  $\rho_{123}$ (\ref{3X})
and $\sigma_{123}$ (\ref{3X-class2}). The algebraic structures of both three-qubit density matrices
offer many simplification in quantifying  geometric quantum discord. To deal with the states
$\rho_{123}$ (\ref{3X}) and $\sigma_{123}$ (\ref{3X-class2}) in a
common framework, it is interesting to note that $\rho_{123}$ as
well as $\sigma_{123}$  have a similar Fano-Bloch representation.
That is
\begin{equation}\label{3X-fano-general}
\varrho_{123}=\frac{1}{8}\bigg[T_{000} ~\sigma_0\otimes
\sigma_0\otimes \sigma_0 + T_{300}~\sigma_3\otimes \sigma_0\otimes
\sigma_0 + \sum_{(\beta,\gamma)\neq (0,0)} T_{0 \beta \gamma}~
\sigma_{0}\otimes \sigma_{\beta}\otimes \sigma_{\gamma} +  \sum_i
\sum _{(\beta,\gamma)\neq (0,0)} T_{i \beta \gamma}
~\sigma_{i}\otimes \sigma_{\beta}\otimes \sigma_{\gamma}\bigg]
\end{equation}
where the notation  $T_{\alpha\beta\gamma}$ stands for the
correlations coefficients  ${\cal R}_{\alpha\beta\gamma}$ (resp.
${\cal T}_{\alpha\beta\gamma}$) of the  states $\varrho_{123}$ of type
$\rho_{123}$
(\ref{3X})(resp. $\sigma_{123}$ (\ref{3X-class2})). The evaluation of the geometric quantum discord (\ref{definition-discord}) requires
a minimization procedure over the set of all classically correlated states,
i.e., the states of the form (\ref{xi12}). In a bipartition of type
$1\vert 23$, a zero discord state is necessarily of the form
\begin{equation}\label{chi}
\chi_{1|23} = p_1 \vert \psi_1 \rangle \langle \psi_1 \vert \otimes
\varrho_1^{23} + p_2 \vert \psi_2 \rangle \langle \psi_2 \vert
\otimes \varrho_2^{23}
\end{equation}
where $\{ \vert \psi_1 \rangle , \vert \psi_2 \rangle \}$ is an
orthonormal basis related to the qubit $1$. The density matrices
$\varrho_i^{23}$ ($i=1,2$) corresponding to the subsystem $23$ write
as
$$ \varrho_i^{23} = \frac{1}{4} \bigg[  \sum_{\alpha,\beta} {\rm Tr}(\varrho_i^{23} \sigma_{\alpha}\otimes \sigma_{\beta})\sigma_{\alpha}\otimes \sigma_{\beta}\bigg].$$
The Fano-Bloch form of  the tripartite classical state (\ref{chi}) is
\begin{equation}\label{chi-matrix}
\chi_{1|23} = \frac{1}{8}\Bigg[ \sigma_{0}\otimes \sigma_{0} \otimes \sigma_{0}
+ \sum_{i=1}^3 te_i ~~\sigma_i \otimes \sigma_{0} \otimes \sigma_{0} $$
$$+ \sum_{(\alpha,\beta)\neq (0,0)} (s_+)_{\alpha,\beta} ~\sigma_0 \otimes \sigma_{\alpha}\otimes \sigma_{\beta}
+ \sum_{i=1}^3  \sum_{(\alpha,\beta)\neq (0,0)}  e_i (s_-)_{\alpha,\beta} ~\sigma_i \otimes \sigma_{\alpha}\otimes \sigma_{\beta}\Bigg]
\end{equation}
where
$$ t = p_1-p_2 \qquad e_i = \langle \psi_1 \vert \sigma_i \vert \psi_1 \rangle  \qquad
(s_{\pm})_{\alpha,\beta} = {\rm Tr}\big((p_1\varrho_1^{23} \pm
p_2\varrho_2^{23}) \sigma_{\alpha}\otimes \sigma_{\beta}\big).$$ The
Hilbert-Schmidt distance between the state $\varrho_{123}$
(\ref{3X-fano-general}) and a classical state of type (\ref{chi-matrix})
gives
\begin{equation}\label{HS0}
|| \varrho_{1|23} - \chi_{1|23}||^2 = \frac{1}{8} \bigg[  (t^2
-2te_3T_{300} + T^2_{300}) +  \sum_{(\alpha,\beta)\neq (0,0)}
(T_{0\alpha\beta} -  (s_+)_{\alpha,\beta})^2 + \sum_{i=1}^3
\sum_{(\alpha,\beta)\neq (0,0)} (T_{i\alpha\beta} -
e_i(s_-)_{\alpha,\beta})^2  \bigg].
\end{equation}
Setting zero the partial derivatives of
Hilbert-Schmidt distance (\ref{HS0}) with respect to the variables
$t$ and $(s_{\pm})_{\alpha,\beta}$,  one has
\begin{equation}\label{solu}
t = e_3 T_{300}  \qquad (s_{+})_{\alpha,\beta} = T_{0\alpha\beta}   \qquad (s_{-})_{\alpha,\beta} = \sum_{i=1}^3 e_i T_{i\alpha\beta}.
\end{equation}
Reporting the results (\ref{solu}) in (\ref{HS0}), one obtains
\begin{equation}\label{HS}
|| \varrho_{1|23} - \chi_{1|23}||^2 = \frac{1}{8} \bigg[  T^2_{300}
- e^2_3T^2_{300} + \sum_{i=1}^3  \sum_{(\alpha,\beta)\neq (0,0)}
T^2_{i\alpha\beta} - \sum_{i,j=1}^3  \sum_{(\alpha,\beta)\neq (0,0)}
e_ie_jT_{i\alpha\beta}T_{j\alpha\beta}\Bigg]
\end{equation}
to be optimized with respect to the three components of the unit vector $\vec{e}~^t = (e_1, e_2, e_3)$.
The equation (\ref{HS}) can re-expressed as
\begin{equation}\label{HS1}
|| \varrho_{1|23} - \chi_{1|23}||^2 = \frac{1}{8} \big[  ||x||^2 +
||T||^2 -\vec{e}~^t(xx^t + TT^t)\vec{e}\big]
\end{equation}
in terms of the   $3\times 1$ matrix $x$ defined by
\begin{equation}\label{x}
x^t=(0,0,T_{300})
\end{equation}
and the $3\times 15$ matrix given by
\begin{equation}\label{T}
T = (T_{i\alpha  \beta}) ~~~ {\rm with}~~~~ i=1,2,3
~~~~(\alpha,\beta) \neq (0,0).
\end{equation}
The minimal value of the Hilbert-Schmidt distance (\ref{HS1}) is
reached when $\vec{e}$ is the eigenvector associated with the
largest eigenvalue $k_{\rm max}$ of the matrix defined by
\begin{equation}\label{matrixK-class2}
K = xx^{t}+TT^{t}.
\end{equation}
It follows that the minimal value given by
\begin{equation}\label{Dg}
D_{\rm g} (\varrho_{1|23}) = \frac{1}{8} (k_1 + k_2 + k_3 - k_{\rm
max})
\end{equation}
is the measure quantifying the pairwise quantum discord in the
state $\varrho_{123}$ divided into  the subsystems $1$ and
$23$. Note that the sum of the eigenvalues $k_1$, $k_2$ and $k_3$ of
the matrix $K$ is exactly the sum of the Hilbert-Schmidt norms of the
matrices $x$ and $T$ $(k_1 + k_2 + k_3  = ||x||^2 + ||T||^2)$. For
the state  $\rho_{123}$ (\ref{3X}) as well as
$\sigma_{123}$ (\ref{3X-class2}),  the matrix $K$ takes the form
\begin{equation}\label{matrixK123}
K=\left(%
\begin{array}{ccc}
K_{11} &  K_{12} & 0 \\
K_{21} & K_{22} & 0 \\
0 & 0 & K_{33} \\
\end{array}%
\right).
\end{equation}
The geometric measure of
quantum discord is determined in terms of the eigenvalues
\begin{equation}\label{k1k2k3}
k_1 =  \frac{1}{2} (K_{11} + K_{22}) + \frac{1}{2} \sqrt {(K_{11} +
K_{22})^2 - 4(K_{11}K_{22} - K_{12}K_{21})}$$
$$k_2 = \frac{1}{2} (K_{11} + K_{22}) - \frac{1}{2} \sqrt {(K_{11} + K_{22})^2 - 4(K_{11}K_{22} - K_{12}K_{21})}$$
$$k_3 = K_{33}.
\end{equation}
Noticing that $k_1$ is always greater that $k_2$, the geometric
quantum discord (\ref{Dg}) rewrites as
\begin{equation}
D_g(\varrho_{1|23}) = \frac{1}{4}~ ( k_2 + {\rm min}(k_1 ,
k_3)).\label{eq:GMQD_new123}
\end{equation}
The minimal Hilbert-Schmidt is obtained for the vector $\vec{e}$
(see equation (\ref{HS1})) associated with the largest eigenvalue of
the matrix K (\ref{matrixK123}). In this sense, to
write the explicit form of closest classical states, one
distinguishes two situations: $k_{\rm max} = k_1$ or $k_{\rm max} =
k_3$. For states $\varrho_{123}$ with entries satisfying the
condition  $k_{\rm max}=k_1$, it is easy to verify that the maximal
eigenvector is given by
$$\vec{e}_1 ^t = (\cos\theta,-\sin\theta,0)  \qquad {\rm with} \qquad \tan \theta =  \frac{K_{11} - k_{1}}{K_{12}},$$
and subsequently the closest classical states write
\begin{equation}\label{chi-matrix-i}
\chi^{(1)}_{1|23} = \frac{1}{8}\Bigg[ \sigma_{0}\otimes \sigma_{0}
\otimes \sigma_{0} + \sum_{(\alpha,\beta)\neq (0,0)}
T_{0\alpha\beta} ~\sigma_0 \otimes \sigma_{\alpha}\otimes
\sigma_{\beta}+ \sum_{(\alpha,\beta)\neq (0,0)}
T^{(1)}_{1\alpha\beta}   ~\sigma_1 \otimes \sigma_{\alpha}\otimes
\sigma_{\beta} + \sum_{(\alpha,\beta)\neq (0,0)}
T^{(1)}_{2\alpha\beta}   ~\sigma_2 \otimes \sigma_{\alpha}\otimes
\sigma_{\beta}\Bigg]
\end{equation}
where
$$ T^{(1)}_{1\alpha\beta} =  \cos^2\theta T_{1\alpha\beta} - \cos\theta \sin\theta T_{2\alpha\beta}
\qquad  T^{(1)}_{2\alpha\beta}  =  \sin^2\theta T_{2\alpha\beta} -
\cos\theta \sin\theta T_{1\alpha\beta}.$$ For states satisfying   $k_{\rm max} = k_3$,
the maximal eigenvector is $$\vec{e}_3 ^t = (0,0,1),$$ and it follows that the closest
classical state takes the form
\begin{equation}\label{chi-matrix-ii}
\chi^{(3)}_{1|23} = \frac{1}{8}\Bigg[ \sigma_{0}\otimes \sigma_{0}
\otimes \sigma_{0} +  T_{300} ~\sigma_3 \otimes \sigma_{0}\otimes
\sigma_{0}+ \sum_{(\alpha,\beta)\neq (0,0)} T_{0\alpha\beta}
~\sigma_0 \otimes \sigma_{\alpha}\otimes \sigma_{\beta} +
\sum_{(\alpha,\beta)\neq (0,0)} T_{3\alpha\beta}   ~\sigma_3 \otimes
\sigma_{\alpha}\otimes \sigma_{\beta}\Bigg].
\end{equation}
It is worth noticing that the entries of the matrix $K$ defined by (\ref{matrixK-class2}) can be explicitly expressed in terms of the correlations factors and subsequently
in terms of the density matrices elements.
Obviously, this will provides us with the analytical expressions of quantum discord
(\ref{eq:GMQD_new123})  in terms  the matrix elements of states
$\rho_{123}$ and $\sigma_{123}$. This issue is discussed in what follows.

\subsubsection{ States of type $\rho_{123}$ }
For three-qubit states $\varrho_{123}$ (\ref{3X-fano-general})
belonging to class of states of type (\ref{3Xfano}), we have
\begin{equation}\label{3X-fano}
\rho_{123}=\frac{1}{8}\bigg[{\cal R}_{000} ~\sigma_0\otimes
\sigma_0\otimes \sigma_0 + {\cal R}_{300}~\sigma_3\otimes
\sigma_0\otimes \sigma_0 +  \sum_{(\beta,\gamma)\neq (0,0)} {\cal
R}_{0 \beta \gamma}~ \sigma_{0}\otimes \sigma_{\beta}\otimes
\sigma_{\gamma} +  \sum_i \sum _{(\beta,\gamma)\neq (0,0)} {\cal
R}_{i \beta \gamma} ~\sigma_{i}\otimes \sigma_{\beta}\otimes
\sigma_{\gamma}\bigg].
\end{equation}
To obtain the matrix $K$ (\ref{matrixK123}), we replace the correlations coefficients
$T_{\alpha\beta \gamma}$, in the matrices $x$ $(\ref{x})$ and $T$
$(\ref{T})$,  with their counterparts ${\cal R}_{\alpha\beta
\gamma}$. In this way, after straightforward algebra, one shows
\begin{equation}\label{elementsK1}
K_{ij} = \sum_{k=1}^{2}\sum_{l=0}^{3}  {\cal R}_{ikl}  {\cal
R}_{jkl} \qquad {\rm with} ~~~i,j = 1,2
\end{equation}
and
\begin{equation}\label{elementsK2}
K_{33} = \sum_{i=0,3}\sum_{j=0}^{3}  {\cal R}^2_{3ij}.
\end{equation}
Furthermore, using the relations (\ref{relation3ret2r}), these
quantities are expressed as
\begin{equation}
K_{11}=2[(R_{11}^{00})^2+(R_{11}^{11})^2]+2[(R_{12}^{00})^2+(R_{12}^{11})^2]
+4 [ R_{11}^{01}R_{11}^{10}+R_{12}^{01}R_{12}^{10}],
\end{equation}
\begin{equation}
K_{22}=2[(R_{21}^{00})^2+(R_{21}^{11})^2]+2[(R_{22}^{00})^2+(R_{22}^{11})^2]
+4 [ R_{21}^{01}R_{21}^{10}+R_{22}^{01}R_{22}^{10}],
\end{equation}
\begin{equation}
K_{33}=2[(R_{30}^{00})^2+(R_{30}^{11})^2]+2[(R_{33}^{00})^2+(R_{33}^{11})^2]
+4 [R_{30}^{01}R_{30}^{10}+R_{33}^{01}R_{33}^{10}],
\end{equation}
\begin{equation}
K_{12} = K_{21}  = 2[R_{11}^{00}R_{21}^{00}+R_{11}^{11}R_{21}^{11}+ R_{12}^{00}R_{22}^{00}+R_{12}^{11}R_{22}^{11}]
 +2  [ R_{11}^{10}R_{21}^{01}+R_{11}^{01}R_{21}^{10}+
 R_{12}^{10}R_{22}^{01}+R_{12}^{01}R_{22}^{10}],
\end{equation}
in terms of two qubit correlation elements related to the two qubit
correlations matrices $\rho^{ij}$  given by (\ref{Rij}). Subsequently,  the
entries of the matrix $K$ (\ref{matrixK123}) are
\begin{equation}\label{K11K22K33}
K_{11} = 8\bigg(\vert \rho_{23} +  \rho_{41} \vert^2 + \vert \rho_{67} +  \rho_{85} \vert^2 + \vert \rho_{36} +  \rho_{18} \vert^2 +\vert \rho_{54} +  \rho_{72} \vert^2   \bigg)$$
$$ K_{22} = 8\bigg(\vert \rho_{23} -  \rho_{41} \vert^2 + \vert \rho_{67} -  \rho_{85} \vert^2 + \vert \rho_{36} -  \rho_{18} \vert^2 +\vert \rho_{54} -  \rho_{72} \vert^2 \bigg) $$
$$ K_{12} = K_{21} = -16\bigg( \vert \rho_{23} \vert \vert \rho_{14} \vert \sin(\gamma_{23}+ \gamma_{14})  + \vert \rho_{58} \vert \vert \rho_{67} \vert \sin(\gamma_{58} + \gamma_{67})
+ \vert \rho_{18} \vert \vert \rho_{36} \vert \sin(\gamma_{18}- \gamma_{36}) + \vert \rho_{27} \vert \vert \rho_{45} \vert \sin(\gamma_{27} -\gamma_{45})\bigg)$$
$$ K_{33} = 4 \bigg( (\rho_{11} - \rho_{33})^2 + (\rho_{22} -\rho_{44})^2 + (\rho_{55} - \rho_{77})^2 + (\rho_{66} - \rho_{88})^2
+  \vert(\rho_{15} - \rho_{37}) + (\rho_{26} - \rho_{48})\vert^2 +  \vert(\rho_{15}- \rho_{37}) - (\rho_{26} - \rho_{48})\vert^2
\bigg)
\end{equation}
where  $\gamma_{ij} = \frac{\rho_{ij}}{\vert \rho_{ij} \vert}$ for $
i < j$. The results (\ref{K11K22K33})  give the explicit forms
of the  matrix elements of  $K$. Clearly, reporting them in (\ref{k1k2k3}), one can
get the explicit expression of  the geometric quantum
discord (\ref{eq:GMQD_new123}) in terms of the density matrix
elements of $\rho_{123}$. In the particular case where the matrix elements $\rho_{ij}$ are all reals, we have  $K_{12} = K_{21} =0$ and the eigenvalues
$k_1$, $k_2$ and $k_3$ (\ref{k1k2k3}) of the matrix $K$ coincide respectively with $K_{11}$, $K_{22}$ and $K_{33}$. In other hand,
if one ignores the qubit 3, the matrix elements (\ref{K11K22K33})  reduces to ones of two qubit $X$ states and it simply verified that one recovers the results (\ref{vp:lambda12}).

\subsubsection{ States of type $\sigma_{123}$}

Similarly, for states  of type $\sigma_{123}$, we write the matrix density (\ref{3X-fano-class2}) as
follows
\begin{equation}\label{3X-fano}
\sigma_{123}=\frac{1}{8}\bigg[{\cal T}_{000} ~\sigma_0\otimes
\sigma_0\otimes \sigma_0 + {\cal T}_{300}~\sigma_3\otimes
\sigma_0\otimes \sigma_0 +  \sum_{(\beta,\gamma)\neq (0,0)} {\cal
T}_{0 \beta \gamma}~ \sigma_{0}\otimes \sigma_{\beta}\otimes
\sigma_{\gamma} +  \sum_i \sum _{(\beta,\gamma)\neq (0,0)} {\cal
T}_{i \beta \gamma} ~\sigma_{i}\otimes \sigma_{\beta}\otimes
\sigma_{\gamma}\bigg]
\end{equation}
Identifying the coefficients  $T_{\alpha\beta \gamma}$ occurring in
(\ref{3X-fano-general}) with ${\cal T}_{\alpha\beta \gamma}$, one obtains the corresponding matrix $K$
(\ref{matrixK123}) whose elements determine the geometric measure of
quantum discord. Explicitly, we have
\begin{equation}\label{Kij-class2}
K_{kl} = \sum_{i=1,2} \sum_{j=0,3} {\cal T}_{kij}{\cal T}_{lij}+ {\cal T}_{kji}{\cal T}_{lji}
\end{equation}
for $k,l= 1,2$, and
\begin{equation}\label{Kij-class2}
K_{33} = \sum_{i=0,3} \sum_{j=0,3} {\cal T}^2_{3ij} +  \sum_{i=1,2} \sum_{j=1,2}   {\cal T}^2_{3ij}.
\end{equation}
They can be rewritten in terms of
the bipartite correlations matrix $T^{ij}$ associated with the two
qubit density matrices $\sigma^{01}$, $\sigma^{01}$, $\sigma^{10}$
and $\sigma^{11}$ given by (\ref{Tii}) and (\ref{Tij}).  Indeed,
using the relations (\ref{relation3ret2r-class2}) and
(\ref{relation3ret2r-class22}), one shows that  the diagonal elements are given by
\begin{equation}
K_{11}=2[(T_{11}^{00})^2+(T_{11}^{11})^2]+2[(T_{12}^{00})^2+(T_{12}^{11})^2]+4|
T^{01}_{10}|^2+4 |T^{01}_{13}|^2
\end{equation}
\begin{equation}
K_{22}=2[(T_{21}^{00})^2+(T_{21}^{11})^2]+2[(T_{22}^{00})^2+(T_{22}^{11})^2]+4|
T^{01}_{20}|^2+4 |T^{01}_{23}|^2
\end{equation}
\begin{equation}
K_{33}= 2[(T_{30}^{00})^2+(T_{30}^{11})^2] +2[(T_{33}^{00})^2+(T_{33}^{11})^2]+4|
T^{01}_{31}|^2+4 |T^{01}_{32}|^2
\end{equation}
where we have used the relation $\overline{T^{01}_{\alpha,\beta}} =
T^{10}_{\alpha,\beta}$. The non zero  off-diagonal element $K_{12}$ rewrites
\begin{equation}
K_{12}= K_{21}=2({T}_{21}^{00}T_{11}^{00}+T_{21}^{11}{T}_{11}^{11})+2({T}_{22}^{00}T_{12}^{00}+T_{22}^{11}{T}_{12}^{11})$$
$$+2(\overline{T^{01}_{20}}T^{01}_{10}+T^{01}_{20}\overline{T^{01}_{10}})
+2(\overline{T^{01}_{23}}T^{01}_{13}+T^{01}_{23}\overline{T^{01}_{13}}).
\end{equation}
Finally,  using the relations (\ref{Tii}) and (\ref{Tij}), one gets
\begin{equation}\label{L11}
K_{11} = 8 \big[  |\sigma_{41} + \sigma_{23}|^2 + |\sigma_{85} +\sigma_{67}|^2 \big] +
4\big[|\sigma_{17}+\sigma_{35} +\sigma_{28}+\sigma_{46}|^2+|\sigma_{17}+\sigma_{35}-\sigma_{28}-\sigma_{46}|^2\big],
\end{equation}
\begin{equation}\label{L22}
K_{22} = 8 \big[  |\sigma_{41} - \sigma_{23}|^2 + |\sigma_{85} -\sigma_{67}|^2 \big] +
4\big[|\sigma_{17}-\sigma_{35} +\sigma_{28}-\sigma_{46}|^2+|\sigma_{17}-\sigma_{35}-\sigma_{28}+\sigma_{46}|^2\big],
\end{equation}
\begin{equation}\label{L33}
K_{33} = 4\bigg[(\sigma_{11}-\sigma_{33})^2+(\sigma_{22}-\sigma_{44})^2+ (\sigma_{55}-\sigma_{77})^2+(\sigma_{66}-\sigma_{88})^2
+ |\sigma_{16}-\sigma_{38}-\sigma_{47}+\sigma_{25}|^2+|\sigma_{16}-\sigma_{38}+\sigma_{47}-\sigma_{25}|^2\bigg],
\end{equation}
and
\begin{equation}\label{L12}
K_{12} =  -16 \bigg[|\sigma_{23}||\sigma_{14}|\sin(\alpha_{23} +
\alpha_{14}) + |\sigma_{58}||\sigma_{67}|\sin(\alpha_{58} +
\alpha_{67}) + |\sigma_{35}||\sigma_{17}|\sin(\alpha_{17} -
\alpha_{35}) + |\sigma_{28}||\sigma_{46}|\sin(\alpha_{28}
-\alpha_{46}) \bigg]
\end{equation}
where  $\alpha_{ij} = \frac{\sigma_{ij}}{\vert \sigma_{ij} \vert}$
for $ i < j$. Substituting the quantities (\ref{L11}), (\ref{L22}), (\ref{L33}) and (\ref{L12}) in the expressions
(\ref{k1k2k3}), we have the  geometric discord
(\ref{eq:GMQD_new123}) in terms of matrix elements of $\sigma_{123}$ (\ref{3X-fano-class2}). In the special situation where all
the entries of the density matrix $\sigma_{123}$ are reals, we have $k_1 = K_{11}$, $k_2 = K_{22}$ and $k_3 = K_{33}$.

\section{Monogamy of geometric discord in three-qubit $X$ states}

The quantum correlation can be transferred between the components of a quantum system comprising many parties. This shareability  is however subject to the monogamy relation which is given for a three-qubit system by
$$ Q_{1|23} \geq Q_{1|2} + Q_{2|3}$$
where $Q$ denotes a measure of pairwise quantum correlation in the system. This inequality means that the amount of quantum correlation shared between the qubits 1 and 2 restricts the possible amount of quantum correlation between the qubits 2 and 3 so that the sum is always less than the total bipartite correlation between the qubit 1 and the subsystem containing the qubits 2 and 3.
This important property was originally proposed  by Coffman, Kundo and Wootters in 2001
\cite{Coffman} for squared concurrence and extended since then to
other correlation quantifiers such as entanglement of formation
\cite{Adesso2,Adesso3}, quantum discord  \cite{Giorgi,Prabhu,Sudha,
Allegra,Ren} and its geometric variant \cite{Bruss}. In particular,
the geometric discord was proven to follow the monogamy property on
all pure three-qubit states. Here, we shall
investigate the distribution among the three qubits in the mixed states of
type $\rho_{123}$ (\ref{3X}) and $\sigma_{123}$ (\ref{3X-class2}).
\subsection{Monogamy conditions}

We consider first the states of type (\ref{3X}). The corresponding
reduced matrices $\rho_{12} = {\rm Tr}_3~\rho_{123}$ and $\rho_{13}=
{\rm Tr}_2~\rho_{123}$ are
\begin{equation}\label{reducedrho12}
\rho_{12}=\left(%
\begin{array}{cccc}
  \rho_{11}+\rho_{55} & 0 & 0 & \rho_{14}+\rho_{58} \\
  0 & \rho_{22}+\rho_{66} & \rho_{23}+\rho_{67} & 0 \\
  0 & \rho_{32}+\rho_{76} & \rho_{33}+\rho_{77} & 0 \\
  \rho_{41}+\rho_{85} & 0 & 0 & \rho_{44}+\rho_{88} \\
\end{array}%
\right)
\end{equation}
\begin{equation}\label{reducedrho13}
\rho_{13}=\left(%
\begin{array}{cccc}
  \rho_{11}+\rho_{22} & \rho_{15}+\rho_{26} & 0 & 0 \\
\rho_{51}+\rho_{62} & \rho_{55}+\rho_{66} & 0 & 0 \\
  0 & 0 & \rho_{33}+\rho_{44} & \rho_{37}+\rho_{48} \\
  0 & 0 & \rho_{73}+\rho_{84} & \rho_{77}+\rho_{88} \\
\end{array}%
\right).
\end{equation}
The reduced two qubit states   $\rho_{12}$ (\ref{reducedrho12}) is
$X$-shaped. The bipartite geometric discord can be derived
using the results (\ref{vp:lambda12}). Therefore, the bipartite quantum
correlation in the state $\rho_{12}$, as measured by Hilbert-Schmidt
distance, is
\begin{equation}\label{QDrho12}
D_g (\rho_{12}) = \frac{1}{4} ( q_2 + {\rm min}(q_1+q_3))
\end{equation}
where
\begin{equation}
q_1=4(\mid\rho_{14}+\rho_{58}\mid+\mid\rho_{23}+\rho_{67}\mid)^2$$
$$ q_2=4(\mid\rho_{14}+\rho_{58}\mid-\mid\rho_{23}+\rho_{67}\mid)^2$$
$$
q_3=2[(\rho_{11}+\rho_{55}-\rho_{33}-\rho_{77})^2+(\rho_{22}+\rho_{66}-\rho_{44}-\rho_{88})^2].
\end{equation}
The state $\rho_{13}$ (\ref{reducedrho13}) is classically correlated. The quantum correlation
between the qubits
 $1$ and $3$ is zero. This is easily verified using the prescription
described previously to get the
 discord in an arbitrary two qubit state. In this special case, the
eigenvalues of the analogue of the matrix $K$ (\ref{matrixK12}) are
 $$p_1 = p_2 = 0$$
$$ p_3 = 2 [(\rho_{11}+\rho_{22} - \rho_{33}- \rho_{44})^2 +
(\rho_{55}+\rho_{66} - \rho_{77}-\rho_{88})^2]$$
which implies that the geometric discord is indeed zero:
\begin{equation}\label{QDrho13}
D_g (\rho_{13}) = 0.
\end{equation}
It follows that the geometric discord in the three-qubit states
$\rho_{123}$ (\ref{3X}) is monogamous  when
\begin{equation}
D_g (\rho_{1\vert 23}) \geq D_g (\rho_{12})
\end{equation}
where $D_g (\rho_{1\vert 23})$ and $D_g (\rho_{12})$ are
respectively given by (\ref{eq:GMQD_new123}) and (\ref{QDrho12}).\\
Analogously, for the states of type $\sigma_{123}$ (\ref{3X-class2}),
the reduced two qubit states  are
\begin{equation}\label{reducedsigma12}
\sigma_{12}=\left(%
\begin{array}{cccc}
  \sigma_{11}+\sigma_{55} & 0 & 0 &\sigma_{14}+\sigma_{58} \\
  0 & \sigma_{22}+\sigma_{66} &\sigma_{23}+\sigma_{67} & 0 \\
  0 & \sigma_{32}+\sigma_{76} & \sigma_{33}+\sigma_{77} & 0 \\
  \sigma_{41}+\sigma_{85} & 0 & 0 & \sigma_{44}+\sigma_{88} \\
\end{array}%
\right)
\end{equation}
\begin{equation}\label{reducedsigma13}
\sigma_{13}=\left(%
\begin{array}{cccc}
  \sigma_{11}+\sigma_{22} & 0 & 0 &\sigma_{17}+\sigma_{28} \\
  0 & \sigma_{55}+\sigma_{66} &\sigma_{53}+\sigma_{64} & 0 \\
  0 & \sigma_{35}+\sigma_{46} & \sigma_{33}+\sigma_{44} & 0 \\
  \sigma_{71}+\sigma_{82} & 0 & 0 & \sigma_{77}+\sigma_{88} \\
\end{array}%
\right).
\end{equation}
The two qubit density matrices $\sigma_{12}$ and $\sigma_{13}$ are
$X$ shaped. It follows that the geometric measure of pairwise
quantum  discord arises directly   from the results
(\ref{vp:lambda12}) modulo the appropriate substitutions. Accordingly,  for $\sigma_{12}$, the geometric
quantum discord is
\begin{equation}\label{QDsigma12}
D_g(\sigma_{12}) = \frac{1}{4} {\rm min}(l_1+l_2, l_3+l_2)
\end{equation}
where
$$l_1 = 4 ( \vert \sigma_{14}+\sigma_{58} \vert + \vert
\sigma_{23}+\sigma_{67} \vert)^2$$
$$ l_2 = 4 ( \vert \sigma_{14}+\sigma_{58} \vert - \vert
\sigma_{23}+\sigma_{67} \vert)^2 $$
$$ l_3 = 2[( \sigma_{11}+\sigma_{55} - \sigma_{33}-\sigma_{77} )^2 + (
\sigma_{22}+\sigma_{66} - \sigma_{44}-\sigma_{88})^2].$$
In the same way, for the subsystem described by $\sigma_{13}$,
one gets
\begin{equation}\label{QDsigma13}
 D_g(\sigma_{12}) = \frac{1}{4} {\rm
min}(m_1+m_2, m_3+m_2)
\end{equation}
where
$$m_1 = 4 ( \vert \sigma_{17}+\sigma_{28}\vert + \vert
\sigma_{53}+\sigma_{64}\vert)^2$$
$$ m_2 = 4 ( \vert \sigma_{17}+\sigma_{28} \vert - \vert
\sigma_{53}+\sigma_{64} \vert)^2$$
$$ m_3 =  2[( \sigma_{11}+\sigma_{22}-\sigma_{33}-\sigma_{44})^2 +
(\sigma_{55}+\sigma_{66}- \sigma_{77}-\sigma_{88} )^2].$$
The geometric discord satisfy the monogamy properly when the entries
of the density matrix $\sigma_{123}$ satisfy the inequality
\begin{equation}
D_g (\sigma_{1\vert 23}) \geq D_g (\sigma_{12}) + D_g (\sigma_{13})
\end{equation}
where the $D_g (\sigma_{1\vert 23})$ is evaluated from
(\ref{eq:GMQD_new123}). To exemplify these results, we consider some
special instances of mixed three-qubit states.

\subsection{Some special mixed states}
\subsubsection{Mixed GHZ-states}
We consider the mixed three-qubit ${\rm GHZ}$ state defined by
\begin{equation}
\rho_{\rm GHZ} = \frac{p}{8}  ~\mathbb{I} + (1-p) ~\vert {\rm GHZ}
\rangle \langle {\rm GHZ}  \vert
\end{equation}
where the pure ${\rm GHZ} $-state is given by
$\vert {\rm GHZ}  \rangle = \frac{1}{\sqrt{2}} (\vert 000 \rangle + \vert
111  \rangle).$
The states $\rho_{\rm GHZ}$ belong to the class of mixed three-qubit
states of type $\rho_{123}$ (\ref{3X}). Subsequently, using the
expressions (\ref{K11K22K33}), it is simple to verify that the
eigenvalues of the matrix $K$  are
$$ \lambda_1 = \lambda_2 = \lambda_3 = 2 (1-p)^2$$
and thus the geometric measure of the pairwise discord between the
subsystems $1$ and $23$ is
\begin{equation}\label{DGGHZ}
D_g (\rho_{\rm GHZ}) = \frac{1}{2} (1-p)^2.
\end{equation}
The maximal value of quantum correlation is reached for $p=0$ (pure
${\rm GHZ}$ state), and for $p=1$ the discord is vanishing as
expected. To discuss the monogamy, we determine the pairwise
geometric discord in the subsystems containing the qubits $1-2$ and
the qubits $1-3$. We denote the associated states by $\rho_{{\rm
GHZ}_{12}}$ and $\rho_{{\rm GHZ}_{23}}$ respectively. Using the
results (\ref{QDrho12}) and (\ref{QDrho13}), one obtains
$$ D_g (\rho_{{\rm GHZ}_{12}}) = 0  \qquad D_g (\rho_{{\rm GHZ}_{13}}) = 0.$$
Using the result (\ref{DGGHZ}), one has
\begin{equation}
D_g (\rho_{\rm GHZ}) \geq  D_g (\rho_{{\rm GHZ}_{12}}) + D_g
(\rho_{{\rm GHZ}_{13}}),
\end{equation}
reflecting that the quantum discord in the states $\rho_{\rm GHZ}$,
as  quantified  by Hilbert-Schmidt norm, follows the monogamy
constraint.
\subsubsection{Mixed W-states}
The second example deals with a special type of three-qubit states
$\sigma_{123}$ (\ref{3X-class2}). They are given by
\begin{equation}
\sigma_{\rm W} = \frac{p}{8}  ~\mathbb{I} + (1-p) ~\vert {\rm W}
\rangle \langle {\rm W}  \vert.
\end{equation}
in terms of the $W$ state:
$\vert {\rm W}  \rangle = \frac{1}{\sqrt{3}} \vert 100 \rangle + \vert 010
 \rangle + \vert 001  \rangle.$
Using the expressions (\ref{L11})-(\ref{L12}), one gets
$$\lambda_1 = \lambda_2 = \frac{16}{9} (1-p)^2 \qquad \lambda_3 =
\frac{20}{9} (1-p)^2$$
and the  geometric discord reads as
\begin{equation}
D_{\rm g} (\sigma_{\rm W})= \frac{4}{9} (1-p)^2.
\end{equation}
In other hand, from the equations (\ref{QDsigma12}) and
(\ref{QDsigma13}), one has
\begin{equation}
D_{\rm g} (\sigma_{{\rm W}_{12}})= D_{\rm g} (\sigma_{{\rm
W}_{13}})= \frac{1}{6} (1-p)^2.
\end{equation}
where $\rho_{{\rm W}_{12}}$ and $\rho_{{\rm W}_{23}}$ are the two
qubit states corresponding to the subsystems comprising the qubits
1-2 and 1-3 respectively. It is clear that
\begin{equation}
D_g (\rho_{\rm W}) \geq  D_g (\rho_{{\rm W}_{12}}) + D_g (\rho_{{\rm
W}_{13}}).
\end{equation}
The geometric measure of quantum discord in the states $\sigma_{\rm W}$
satisfies the monogamy condition.
\subsubsection{Three-qubit state of Bell type}
Finally, we consider the three-qubit
\begin{equation}
\rho_{\rm B} = \frac{1}{8}  \bigg( \sigma_0\otimes\sigma_0\otimes\sigma_0
+ \sum_{i=0}^3  c_i \sigma_i\otimes\sigma_i\otimes\sigma_i\bigg).
\end{equation}
which can be viewed as the extended version of two-qubit Bell state. The
state $\rho_{\rm B}$
has non vanishing matrix elements only along the diagonal and off
diagonal. Indeed,
in the computational basis, it writes
\begin{equation}\label{Bell}
\rho_{\rm B}=\frac{1}{8}\left(%
\begin{array}{cccccccc}
1+c_3 & 0 & 0 & 0 & 0 & 0 & 0 & c_1+ic_2 \\
  0 & 1-c_3  & 0 & 0 & 0  & 0 & c_1-ic_2& 0 \\
  0 & 0 & 1-c_3  & 0 & 0 & c_1-ic_2 & 0 & 0 \\
  0 & 0 & 0 & 1+c_3  &  c_1+ic_2 & 0 & 0 &  0 \\
  0 & 0 & 0 & c_1-ic_2 & 1-c_3  & 0 & 0 & 0 \\
 0 &  0& c_1+ic_2 & 0 & 0 & 1+c_3  & 0 & 0 \\
  0 & c_1+ic_2& 0& 0 & 0 & 0 & 1+c_3  & 0 \\
c_1-ic_2 & 0 & 0 & 0 & 0 & 0 & 0 &1+c_3  \\
\end{array}%
\right).
\end{equation}
From equations (\ref{K11K22K33}), one has
$$ K_{11}  = c^2_1 \qquad K_{22} = c^2_2 \qquad K_{33} = c^2_3 \qquad
K_{21} = K_{21} = 0$$
and the geometric discord in the bipartition  $1\vert 23$ is
\begin{equation}
D_{\rm g}( \rho_{B}) = \frac{1}{8} (c^2_1 + c^2_2 + c^2_3 - c^2)
\end{equation}
where $c^2 = {\rm max}( c^2_1 , c^2_2 , c^2_3 ) $. It is remarkable
that the reduced two qubit states  given by
$$ \rho_{B_{12}} = \rho_{B_{13}} = \frac{1}{4} \sigma_0\otimes\sigma_0$$
do not present quantum correlations when  measured by the
Hilbert-Schmidt distance $(i.e., D_{\rm g} (\rho_{B_{12}} ) = D_{\rm
g} (\rho_{B_{13}} ) = 0)$. The quantum  $D_{\rm g}( \rho_{B})$ is
always non negative and therefore the geometric discord in the
states $\rho_{\rm B}$ is monogamous.

\section{Concluding remarks}

In this work, we have investigated the analytical derivation  of
quantum correlations in mixed states describing quantum systems
comprising three qubits.  We have deliberately considered the square norm
(Hilbert-Schmidt
distance) instead of entropic based quantifiers. In fact, despite the
information meaning of based entropy
measures, determining  explicit expressions of quantum correlations
requires optimization procedures that are in general very
complicated to achieve even in two qubit systems. In this respect,
the geometric quantifiers are advantageous in obtaining closed
computable expressions of the information contained in a tripartite
quantum system.  In this picture, through the geometrized variant of
quantum discord, we characterized the bipartite quantum correlations in
mixed three-qubit states and their analytic expressions  are
explicitly derived  for two families of generalized three-qubit $X$-
states.   In addition,  we have
determined the explicit Fano-Bloch expressions of classically correlated
(zero
discord) states. In other hand, we  have studied the monogamy property
and the shareability
limitations of geometric quantum discord
for  two kinds  of generalized three-qubit $X$ states. To exemplify our results, we discussed the monogamy
property in mixed three-qubit states of ${\rm W}$, ${\rm GHZ}$ and
Bell types.\\

\noindent Finally, it worth to notice that the geometric
measure of quantum discord obtained in this paper are  useful for many
purposes. First, it provides
the explicit amount of quantum correlation in mixed three-qubit $X$
states that is generalizable to arbitrary quantum systems of
arbitrary number of qubits. Also, it offers a computable tool to
get the multipartite quantum correlation defined as the sum of all
pairwise partition in a multi-components system (see for instance
\cite{daoud2}). In this sense, the present approach constitutes a good
alternative to evaluate tripartite correlation in mixed states
generalizing the analysis done for pure tripartite systems
\cite{Bruss,daoud2,daoud3}. In other hand, this approach  is ready
to adapt in investigating the dynamics of geometric discord in
quantum systems subjected to decoherence mechanisms in the spirit of
the results recently presented  in \cite{J.Zhou}. Further study in
this direction might be worthwhile.

\end{document}